\documentclass[a4paper,fleqn,usenatbib]{mnras}

\usepackage{graphicx}   
\usepackage{amsmath}    
\usepackage{amssymb}    

\title[]{Secular evolution of close--in planets: the effects of general relativity}

\author[F. Marzari and M. Nagasawa]{
F. Marzari,$^{1}$\thanks{E-mail: marzari@pd.infn.it}
M. Nagasawa,$^{2}$
\\
$^{1}$Department of Physics and Astronomy, University of Padova, Via Marzolo 8, 35131 Padova, Italy\\
$^{2}$ Kurume University, School of Medicine, Department of Physics, 67 Asahi--machi, Kurume--city, Fukuoca, 830--0011, Japan. 
}

\date{.....}
\pubyear{2020}

\begin{document}
\label{firstpage}
\pagerange{\pageref{firstpage}--\pageref{lastpage}}
\maketitle

\begin{abstract}
Pairs of planets in a system may end up close to their host star on eccentric orbits as a consequence of planet--planet scattering, Kozai or secular migration. 
In this scenario,  general relativity and secular perturbations have comparable timescales and may interfere with each other with relevant effects on the eccentricity and pericenter evolution of the two planets. We explore, both analytically and via numerical integration, how the secular evolution is changed by general relativity for a wide range of different initial conditions.  We find that when the faster secular frequency approaches the general relativity precession rate, which tipically occurs when the outer planet moves away from the inner one, it relaxes to it and a significant damping of the proper eccentricity of the inner planet occurs. The proper eccentricity of the outer planet is reduced as well  due to the changes in the secular interaction of the bodies. The lowering of 
the peak eccentricities of the two planets during their secular evolution
has important implications on their stability. 
A significant number of two--planet systems, otherwise chaotic because of the mutual secular perturbations, are found stable when general relativity is included.

\end{abstract}

\begin{keywords}
{ planets and satellites: general -- planets and satellites: dynamical evolution and stability}
\end{keywords}

\section{Introduction}

A significant fraction of Hot/Warm Jupiters, giant planets with short orbital periods, have high eccentricities. Values as high as 0.6 are observed among known  exoplanets orbiting within 0.2 au from their star. Depending on the strength of the tidal interaction with the star, this may either be a permanent dynamical state (weak tidal interaction) or a transient state during the tidal circularization of the orbit. The timescale for such circularization is highly debated and it depends on the physical properties of both the star and planet but it may take from a few Myrs to Gyrs. 

These Hot/Warm Jupiters may not be lonely bodies and additional planets may be present on outer orbits. This is expected in particular if their high eccentricity is due to planet--planet scattering \citep{weiden-marza96,rasio-ford96,marweiden2002,namouni07,CHATTERJEE08,juritre08,ray08,naga08,ray09,marzari10,naga11,beauge2012,marzagawa2019} possibly triggered by a period of orbital migration by interaction with the protostellar disk. Convergent migration \citep{masset2001,Leepeale2002}, either due to different planet masses and then different migration speeds or caused by the slowing down of the inner planet migration when close to the star or due to the formation of an inner disk hole, may bring the planets into close and unstable orbits which inevitably lead to gravitational encounters.  As shown by \cite{marzagawa2019}, systems of three unstable planets, after a period of planet--planet scattering ending with the ejection of one body, 
well reproduce the observed orbital distribution of the known Hot/Warm Jupiters.  

When more than one massive body is present in the system, the long term gravitational perturbations between the planets lead to a secular evolution characterized by significant oscillations of their eccentricity and inclinations and circulations/librations of pericenter and node longitudes.  However, the perihelion precession induced by General Relativity (hereinafter GR) may significantly alter the secular evolution of two (or more planets). 

\cite{Adams2006} investigated the effects of GR on 3--planet systems like $\Gamma And$ and HD 160691 showing that it could move the systems in and out of secular resonances and potentially increase their stability.  These two systems could also be used as a test of general relativity or, reversing the approach, the GR effects could be used to constraint the orbital parameters of the systems. 
\cite{miga2009} developed a secular theory of high order (up to 24) where they include the effects of GR and the effect of the quadrupole moment of the star.  On the basis of their theory they suggest that GR may play an important role in the secular evolution but they only consider a few test cases. \

\cite{veras2010} have investigated the dynammical evolution of 5 extrasolar planetary systems finding that GR can affect the secular dynamics at high mutual inclinations where the Kozai mechanism may be partly suppressed by GR.  
\cite{zhang2013} focus their study on the eccentricity damping of pair of planets, of which one is close to the star, due to static tides. They include the effects of GR but they are mostly interested in the evolution of low eccentricity planets while they are damped by tides. \cite{marzagawa2019}, while exploring the dynamics of three planets undergoing planet--planet scattering close to their star, found that the surviving two--planets systems have their secular dynamics significantly affected by GR.  

In most previous papers, GR was considered as an ancillary effect and its most important contribution was considered that of inhibiting Kozai librations. In this paper we perform a more systematic study  of  how  GR affects the secular dynamics of two planets and it is able to stabilize systems with initial high eccentricity. This stabilization is not related to the Kozai dynamics but to the interplay between GR and secular perturbations. We perform an extended sampling of the initial conditions of 2 planet systems and explore the evolution of their proper and forced eccentricities and of their main secular frequencies with and without the GR contribution.  We neglect additional perturbing effects like stellar tides or stellar oblatness in order to better outline the GR effects and also because they tipically act on a longer timescale and they may mask the real dynamical effect of GR.

To find the configurations where GR is more effective, we have adopted as a guide  the  Lagrange--Laplace secular theory \citep{murray-dermottSS} (hereinafter LL theory) which has an easy formulation and solution and within which the GR contribution can be easily implemented. In this way we can test the changes in the proper and forced eccentricities of the inner and outer planets and how they change for different initial values of the initial eccentricities.   The LL theory, being limited to terms in $e^2$ and $i^2$, is not very accurate for high values of these orbital parameters. Higher order theories, like that described in  \cite{henrard2005}  or \cite{miga2009} would be best suited  for eccentric and misaligned exoplanets. However, the great complexity of these higher order theories play in favor of a numerical approach, which is adopted here, whose results are interpreted on the basis of the simpler LL theory. 

In Section 2) we show how the predictions of the LL theory change in presence of the additional GR precession term acting mainly on the inner planet. We focus on a coplanar scenario and neglect the evolution of the inclinations.
In Section 3 we describe the numerical model adopted to perform 3--body numerical simulations and also illustrate the outcome of some test model  with and without GR.  In Section 4) we describe  the stabilizing effect of the GR term on the long term evolution of highly eccentric planets. 
In Section 5)  we detail the outcome of a large number of numerical simulations, performed with and without GR, considering two different initial planetary architectures in terms of semi--major axis values. The dependence on the mass of the planets is explored in Section 6. 
Finally, in Section 7 we discuss the results and their implications.

\section{Linear secular theory}

To have an initial glimpse at the effects of GR on the secular evolution of 
two planets close to the central star we  resort to the secular theory of LL \citep{murray-dermottSS}. It can be exploited only for low eccentricities of the planets since it neglects all terms in the disturbing function and in the Lagrange equations higher than second order in eccentricity (and inclination).  This is clearly a limitation when dealing with extrasolar planetary systems where the eccentricities, even very close to the star, can be very high. However, as we will see, the linear theory highlights some of the major effects of GR on the secular evolution of two planets and it gives a useful frame within which to interpret the outcomes of the direct numerical integration of the motion equation. 

\subsection{Brief summary of the LL theory}

The Laplace--Lagrange theory is an averaged secular theory based on an expansion of the disturbing function of the planetary gravitational perturbations truncated to the second order in eccentricity and inclination. It has been developed to compute the long term evolution of the osculating orbital elements of systems of N massive bodies moving around a central more massive one (in our case the star). Once introduced in the Lagrange differential equations to compute the time evolution of the orbital elements, potential singularities may appear and for this reason the non--singular variables $h,k$ and $p,q$ for each planet are adopted. They are defined as: 

\begin{equation} 
\label{eq1}
\begin{split}
h_i & =  e_i \sin (\varpi_i)  \\
k_i & =  e_i \cos (\varpi_i) \\
p_i & =  i_i \sin (\Omega_i) \\
q_i & =  i_i \cos (\Omega_i)  
\end{split}
\end{equation}
 where $e_i$ is the eccentricity of planet $i$ and $\varpi_i$ its pericenter longitude. Similarly, $i_i$ is the inclination and 
 $\Omega_i$ the node longitude of the body, respectively. 
The most interesting feature of the LL theory is that the solution of its equations is simple and 
the time evolution of the non--singular variables is  determined by the solution of a system of first order differential equations with constant terms grouped in the  matrix A for $h,k$ and the  matrix B for $p,q$ (see pg. 276 of \cite{murray-dermottSS} for the analytical definition of the matrices). 
These equations are: 

\begin{equation} \label{eq4}
\begin{split}
 \dot {h_i} &  = k_i  A_{ii}  + k_j A_{ij}  \\
  \dot {k_i} &  = -h_i  A_{ii}  - h_j A_{ij}  \\
   \dot {p_i} &  = q_i  B_{ii}  + q_j B_{ij}  \\
  \dot {q_i} &  = -p_i  B_{ii}  - p_j B_{ij}  \\
\end{split}
\end{equation}
which are first order differential equations with constant coefficients. In fact,
both matrices only depend on the masses and semi--major axes of the two planets. The analytic solution of these equations is given as a combination of trigonometric functions that, in the case of two planets, is simply: 

\begin{equation} \label{eq2}
\begin{split}
h_i & =  e_{i1} \sin(g_1 t + \beta_1) + e_{i2} \sin(g_2 t + \beta_2) \\
k_i & =  e_{i1} \cos(g_1 t + \beta_1) + e_{i2} \cos(g_2 t + \beta_2) \\
p_i & =  i_{i1} \sin( \gamma_1) + i_{i2} \sin(f_2 t + \gamma_2) \\
q_i & =  i_{i1} \cos( \gamma_1) + i_{i2} \cos(f_2 t + \gamma_2) \\
\end{split}
\end{equation}

The fundamental frequencies $g_1$ and $g_2$ are the eigenvalues of the matrix A while the vectors $ e_{i1} $ and $ e_{i2} $ are its eigenvectors. For the evolution of the variables $p,q$ only one frequency, $f_2$, is different from 0 while $f_1 = 0$,  for the conservation of the angular momentum, and $ i_{i1} $ and $ i_{i2} $ are the eigenvectors of the matrix B. The constants $\beta_i$ and $\gamma_i$ and the magnitude of the eigenvectors are determined by the initial conditions (orbital elements)  of the system.
Hereinafter, we will focus on the evolution of the eccentricity and pericenter longitude of the planets, neglecting  the inclination and nodal longitude since we expect that GR mainly affects the former variables through the forced additional pericenter precession. The evolution of the eccentricity of each planet $i$  with time can be computed  by combining the variable $h_i,k_i$: 

\begin{equation} 
\label{eq3}
\begin{split}
e_i(t) & = \sqrt{(h_i(t)^2 + k_i(t)^2)} \\
& = \sqrt{(e_{i1}^2 + e_{i2}^2) + 2 e_{i1} e_{i2} \cos((g_1-g_2)t + \beta_1 - \beta_2)}
\end{split}
\end{equation}

\subsection{Contribution from GR}

When we include the effects  of GR in the secular evolution, 
the matrix A must be updated to include an additional term which strongly depends on the semi--major axis of the planet and its eccentricity. This new term is added to the diagonal elements of the matrix which become  $A_{ii}^{\prime}=A_{ii}+A_{i, {\rm GR}}$  where $A_{i,{\rm GR}}$ is related to the apsidal precession due to GR on planet i,
\begin{equation}
\label{eq6}
A_{i,{\rm GR}}=\frac{3Gn_i M_{\ast}}{a_{pi}c^2(1-e^2_{pi})},
\end{equation}
where $c$ is the speed of light and $n_i$ the mean motion of planet i.

From a first analysis of the contribution of this additional diagonal term due to GR, we expect that in most systems the eccentricity oscillations of the two planets will be  reduced in amplitude. This because the condition for the damping of the eccentricity oscillations is equivalent to the condition that the period of such oscillations becomes small. 
We can see that this corresponds to increasing $g_1-g_2$ which, according to eq.\ref{eq3},  determines the frequency of the sinusoidal oscillations of the eccentricity.
Translating this condition in the language of the matrix A, the amplitude of the eccentricity oscillations with GR is smaller than the amplitude without GR when $A_{11}^{\prime}-A_{22}^{\prime}> A_{11}-A_{22}$, i.e., 
\begin{equation}
\label{eq5}
\frac{A_{1, {\rm GR}}-A_{2, {\rm GR}}}{A_{11}-A_{22}}>0.
\end{equation}
The diagonal elements $A_{ii}$ are related to the apsidal precession forced by the secular perturbations between the two planets (without GR) and the difference  $A_{11}-A_{22}$ is positive in systems with $a_1 m_1^2 < a_2 m_2^2$, which are the cases studied in this paper. 
In addition, the effect of GR is stronger for the inner planet so $A_{1, {\rm GR}}> A_{2, {\rm GR}}$.
Therefore,  in the case of similar mass planets, the apsidal precession of the inner planet, that was originally faster in absence of GR, becomes even faster with GR. This leads to a decrease in periods and amplitudes of the eccentricity oscillations of both the inner and outer planet.

We can prove the predictions of this analytical criterion  by directly solving the secular equations with the additional terms due to the relativistic precession. 
For a pair of Jovian mass planets, $A_{1, {\rm GR}}$ becomes comparable to $A_{11}$ near the star. On the other hand, since the outer planet is usually farther out and not significantly affected by GR, $A_{2, {\rm GR}}$ can be neglected. 
To compute the secular evolution of the two bodies we have then to find the solution of the following differential  equations for $h_i$:
\begin{equation} \label{eq4}
\begin{split}
 \dot {h_1} &  = k_1   \left( A_{11} + \frac {3 G n_1 M_*} {a_{p1} c^2 (1  -e_{p1}^2)} \right ) + k_2 A_{12}  \\
\dot {h_2}  & = k_1   A_{21} +  k_2 A_{22} 
\end{split}
\end{equation}
They cannot be directly solved because of the dependence on $e_{p1}$ which should then be substituted by its expression as a function of $h,k$, significantly complicating the equations. However, as a first approximation, we can adopt  for $e_{p1}$ a constant value given  by the average eccentricity during the secular evolution. This would require an iterative procedure which will converge quickly. Once the GR term is computed with an average value of  $e_{pi}$, the solution would be similar to that described by equations eq.~\ref{eq2} for $h,k$ and eq.~\ref{eq3} for the planetary eccentricity. However, the eigenvalues, i.e. the frequencies $g_1$ and $g_2$, and the eigenvectors, determining the final eccentricity value, will be different due to the change in the diagonal element of the matrix A. According to the previous analytical criteria, this would lead to a damping of the eccentricity oscillations.

\subsection{Solutions with and without GR}

\begin{figure}
\includegraphics[width=\columnwidth]{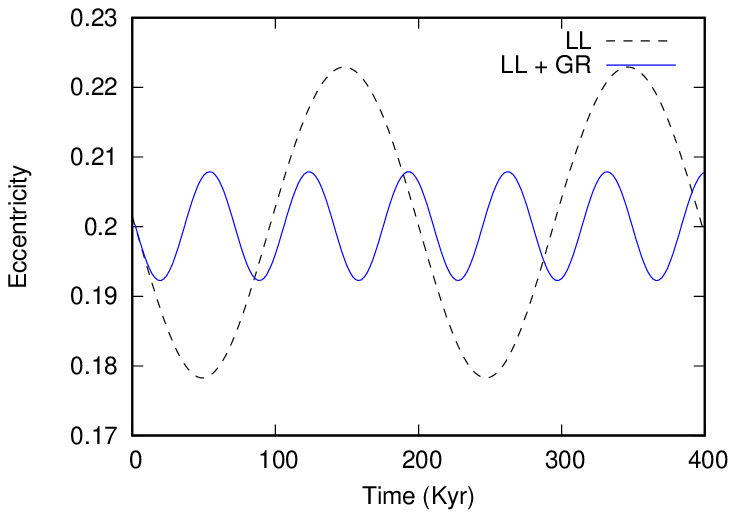}
\includegraphics[width=\columnwidth]{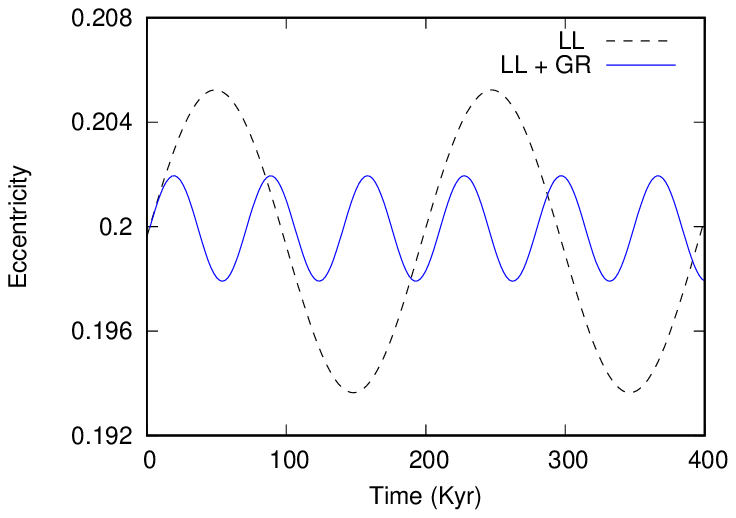}
\caption{\label{fig1} Secular evolution of the inner planet (top plot) and outer planet (bottom plot) with and without the contribution from GR computed analytically from the LL theory. Note the different scale of the two plots with the oscillations of the inner planet wider respect to that of the outer one.
}
\end{figure}

In Fig.~\ref{fig1} we compare the secular evolution predicted with the LL theory with and without the GR term for two planets having an initial semi--major axis of $a_1 =0.1$  and $a_2 =1.5$  au, eccentricity $e_1 = 0.2$ and $e_2 = 0.2$.  The mass of the star is set to $1 M_{\odot}$, value that we will use  throughout this paper, while the initial  pericenter longitudes of the planets are $\pi$ (inner planet) and $\pi/2$ (outer planet). 

 The eccentricity evolution shown in Fig.~\ref{fig1} is computed with eq.~\ref{eq3} which may be simplified as: 

 \begin{equation} \label{eq7}
\begin{split}
e_i(t) & = \sqrt{e_{ai}^2 + e_{bi}^2 \cos((g_1-g_2)t + \beta_1 - \beta_2)}
\end{split}
\end{equation}

where $e_{ai} = \sqrt {e_{i1}^2 + e_{i2}^2}$ while $e_{bi} = \sqrt {2 e_{i1} e_{i2}}$.  

\begin{figure}
\includegraphics[width=\columnwidth]{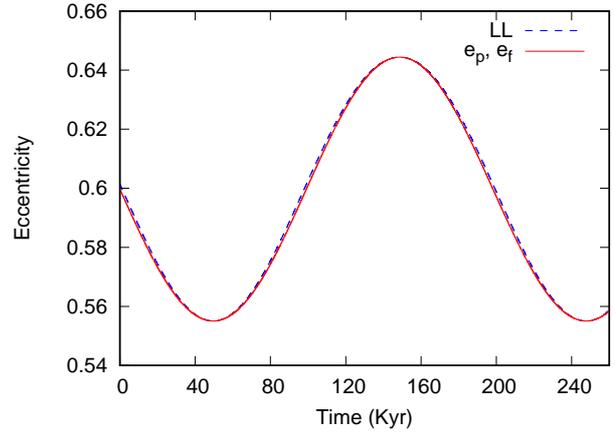}
\caption{\label{fig_compa}
Secular evolution of the eccentricity of a planet computed with the LL theory (eq.~\ref{eq7})  and with the approximated formula given in eq.~\ref{eq8}.  Only minor differences are observed between the predictions of the two equations. 
}
\end{figure}

In this formulation, however, the values of $e_a$ and $e_b$  do not closely recall the observed values and need some interpretation.  In addition, the solutions for higher values of eccentricity are periodic but may depart from the shape predicted by eq.~\ref{eq3}. This because the secular theory is a good approximation only 
when the  eccentricity of the planets is not too high (it is a linear theory). For this reason, we have adopted a rougher fitting to the eccentricity behaviour  which allows to grasp more easily the dynamical behaviour. The eccentricity evolution is simplified and given by:

\begin{equation} \label{eq8}
\begin{split}
e(t) & = e_f + e_p cos((g_1-g_2)t + \beta)
\end{split}
\end{equation}

where $e_f$ may be interpreted as a forced eccentricity, $e_p$ as a proper one while $\beta$ is the integration constant. The values of $e_f$ and $e_p$ are derived from the maximum and minimum values assumed by $e(t)$ during its evolution.   
This choice does not introduce a further approximation to the theory since the predicted evolution of eq.~\ref{eq7} and that of eq.~\ref{eq8} are  almost indistinguishable one from the other, as shown in Fig.\ref{fig_compa}.

By comparing the eccentricity evolution with and without GR in  Fig.~\ref{fig1},
a significant reduction in the oscillation amplitude is observed when the GR term is included, as expected on the basis of the analytical study presented in Sect. 2.1. 
In the formalism of proper and forced eccentricity,  the GR term leads to a smaller $e_{p1}$ while leaving  $e_{f1}$ unchanged. This is related to the increase in the  difference between the two eigenfrequencies.  $g_1$ is significantly higher than $g_2$ when GR is included  changing from 
$2.43 \times 10^{-3}$ deg/yr to $8.82 \times 10^{-3}$ deg/yr while $g_2$, the slower one, is only slightly modified. As a consequence $g_1$ dominates and drives a faster (and damped) oscillation of the eccentricity of both planets. An increase of the initial eccentricity causes a growth of the relativistic precession term which depends on the eccentricity of the inner planet via $1/ (1 - e_p^2)$ and this leads to an even faster $g_1$.  

\begin{figure}
\includegraphics[width=\columnwidth]{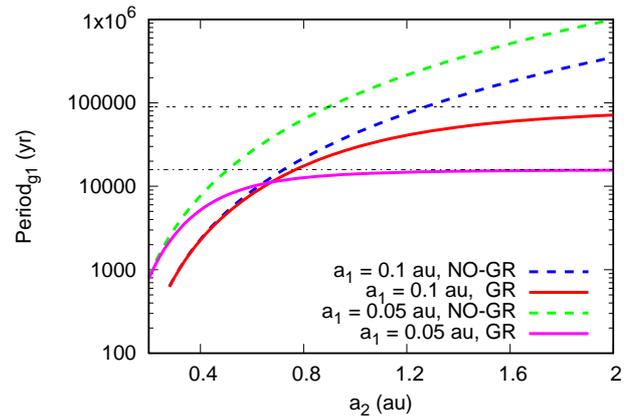}
\caption{\label{fig21} 
Values of the period of $g_1$ as a function of the
outer planet semi--major axis $a_2$ computed with the classical LL theory (dashed lines) and after the inclusion of the GR term (continuous lines)  for $a_1 = 0.1$ au  and $a_1=0.05$ au. The initial values of the planet eccentricities are $e_1 = 0.4$ and $e_2 = 0.4$. The horizontal dotted lines give the period of pericenter precession predicted by the GR term only in the two cases (bottom $a_1=0.05$ au, top $a_1= 0.1$ au). 
}
\end{figure}

\begin{figure}
\includegraphics[width=\columnwidth]{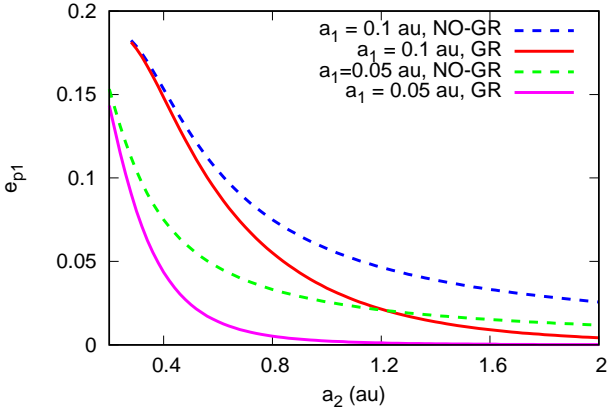}
\caption{\label{fig11}
Predicted values of the eccentricity $e_{p1}$ vs. $a_2$ for the same initial conditions considered in Fig.\ref{fig21} ($a_1 = 0.1$ au and $a_1=0.05$ au). 
The curves for $e_{p1}$ without the contribution from GR (dashed lines) has higher values compared to the case with GR, as expected. The difference is correlated to the growth of $g_1$ when the GR precession is comparable to the secular precession. 
}
\end{figure}

\begin{figure}
\includegraphics[width=0.95\columnwidth]{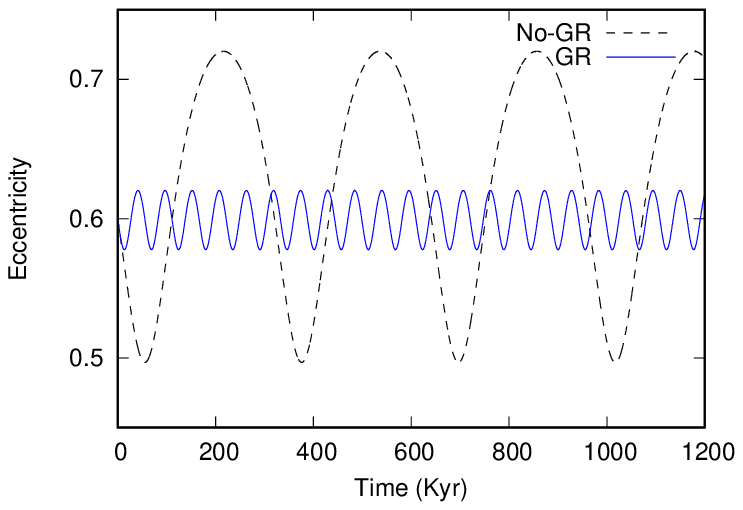}
\vskip -0.2 truecm
\includegraphics[width=0.95\columnwidth]{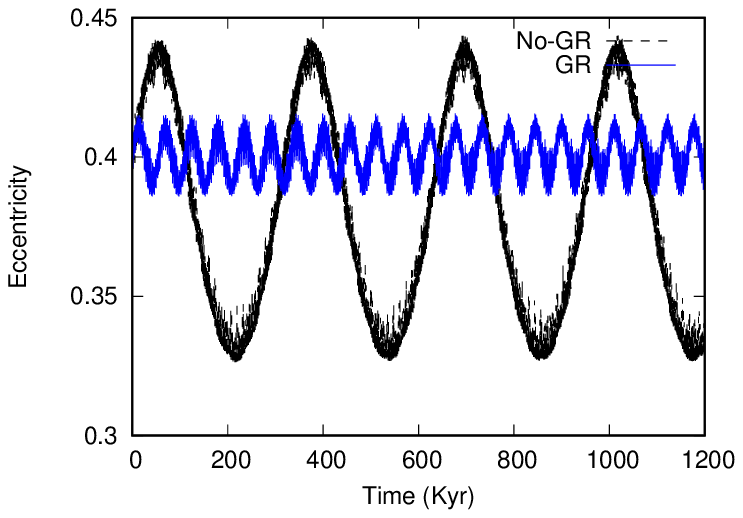}
\vskip -0.2 truecm
\includegraphics[width=0.95\columnwidth]{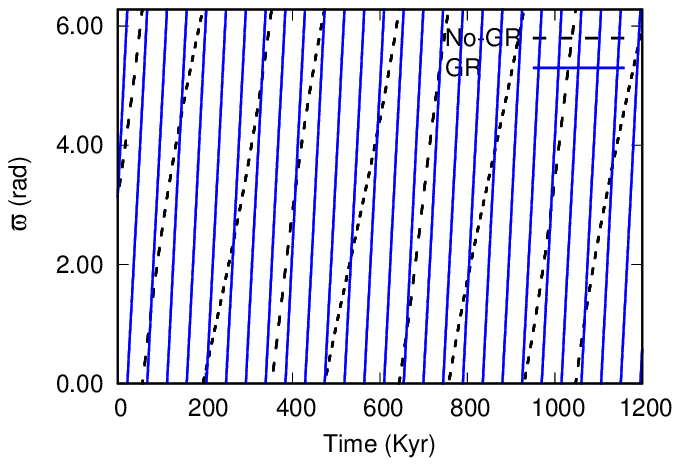}
\vskip -0.2 truecm
\includegraphics[width=0.95\columnwidth]{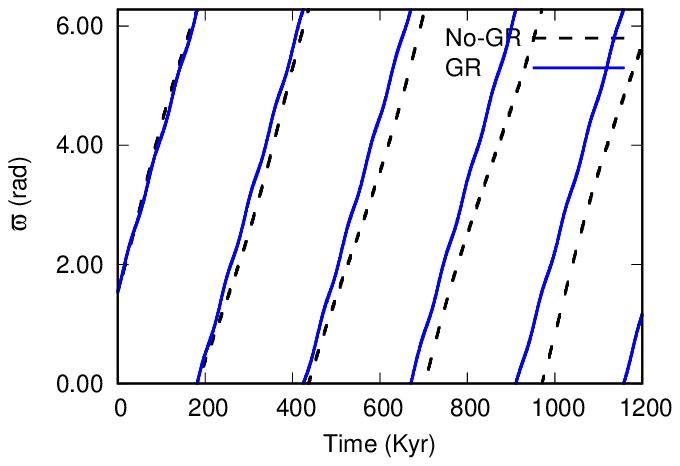}
\vskip -0.2 truecm
\caption{\label{fig2} Secular evolution of the eccentricity of inner planet (top plot) and outer planet (bottom plot) whose orbits have been numerical integrated with and without the contribution from GR. In the two bottom plots the evolution of the pericenter longitude of the two planets is illustrated.
}
\end{figure}

Even if the LL theory is approximated, we can use it to have a first glimpse at the dependence of the planet eccentricity on the semi--major axis of the  outer planet.  In Fig.\ref{fig21} we show the predicted values of the period of the frequency $g_1$  in two different models. In the first one the inner planet has a semi--major axis of $a_1 = 0.1$ au and an eccentricity of $e_1=0.4$ while the outer one has an eccentricity fixed to $e_2=0.4$ and its semi--major axis is regularly sampled  in between 0.2 and 2 au. In the second model, the semi--major axis of the inner planet is decreased to $a_1=0.05$ au while keeping the same its eccentricity and  $a_2$ is  varied as in the previous model.  The two cases without the GR term show an increasing trend to longer periods. When the GR term is added to the LL equations, the period of $g_1$ grows until it saturates at about the value of the relativistic precession period.  In the $A_{11}$ element of the secular matrix, the GR term becomes dominant and one eigenvalue, the frequency $g_1$, approaches the GR precession rate. We expect that when this happens, even the values of eccentricity are significantly effected and damped. This indeed occurs, as shown in 
 Fig.\ref{fig11}.  When the two planets are close, the LL theory predicts only a tiny difference between the values of $e_{p1}$ with and without GR because the secular frequencies are high and the forced precession due to GR is comparatively small and it has little effect on $g_1$ and then on the difference between $g_1$ and $g_2$ which determines the damping of $e_{pi}$. When the outer planet gets farther away from the inner one and $g_1$ approaches the relativistic precession rate, the difference becomes significant and the oscillation amplitude of the eccentricity is damped, as already observed in Fig.\ref{fig1} and 
Fig.\ref{fig2} (outcome of a numerical integration). 

This behaviour suggest us to select for the numerical simulations initial conditions close to the transition from secular--dominated to relativity--dominated systems. In the forthcoming section we will numerically model systems close to this transition.

\section{Numerical modeling}

As anticipated in the previous sections, the LL theory properly works only for small values of the planet eccentricity. As a consequence, to explore the secular evolution of two planets for high values of $e_p$ we numerically integrate the equations of motion with and without the effects of GR, using a fourth-order time-symmetric Hermite code \citep{kokubo98}.

To analyse the numerical results we adopt the model based on the LL theory and fit the eccentricity evolution of each planet with a proper $e_p$ and forced $e_f$ eccentricity (see eq.~\ref{eq8}).
Their values are computed from the periodic evolution of the eccentricity by computing the maximum and minimum values.

The fundamental frequencies $g_1$ and $g_2$ are instead estimated from the evolution of the pericenter longitude $\varpi$ of the two planets. The frequency  values are extracted with an FFT analysis of 
$\varpi_1$ and $\varpi_2$ where $g_1$ and $g_2$ are identified as the two main frequencies dominating the time evolution of these angular variables.

In Fig.\ref{fig2} we show the evolution with time of two planets, the inner one with $a_1 = 0.1$ au and  $e_1 = 0.6$, and the outer one with $a_2 = 1.5$ au and $e_2 = 0.4$. Their pericenters are  initially set to $\pi$ (inner planet) and $\pi/2$ (outer planet).  The numerical integration confirms the reduction of $e_p$, when GR is included, both for the inner and outer planets.  Short term oscillations appear in the eccentricity of the outer planet due to higher order terms which are not included in the LL theory. 
In the bottom panels of  Fig.\ref{fig2} the evolution of the pericenter is shown for both planets. The circulation frequency of $\varpi$ of the inner planet is  significantly faster when the GR term is included, as expected, while  that of the outer planet is only slightly faster with GR. It is interesting to note that the eccentricity of the outer planet is indirectly  affected by GR through the secular perturbations of the inner planet. 

\begin{figure}
\includegraphics[width=\columnwidth]{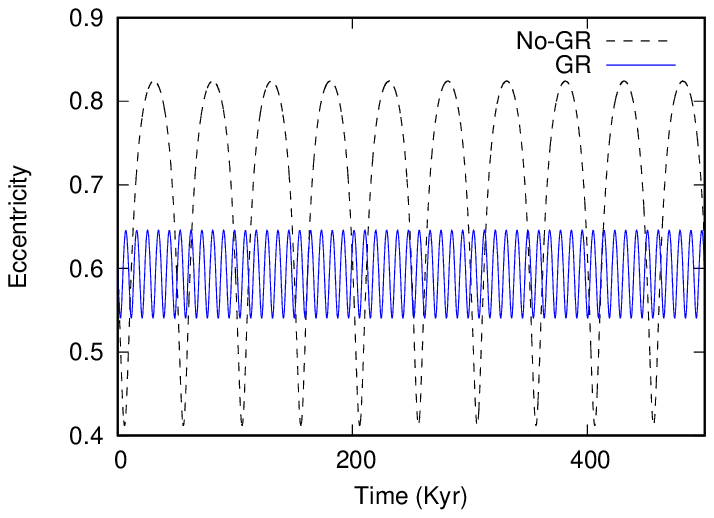}
\includegraphics[width=\columnwidth]{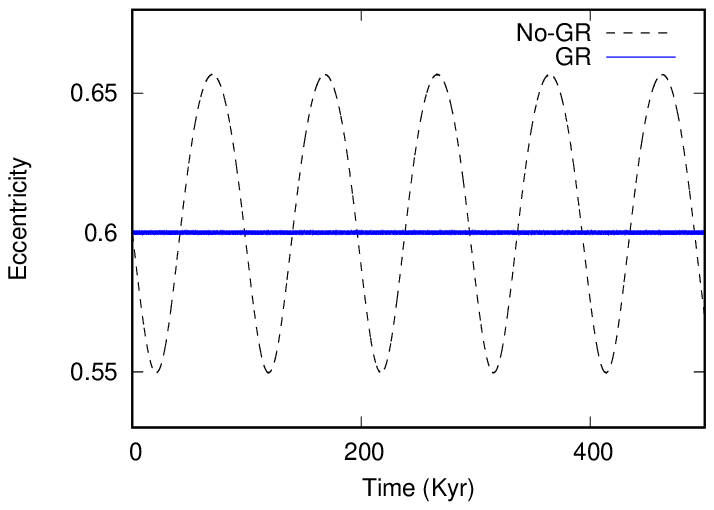}
\caption{\label{fig3} Secular evolution of the inner planet with GR (continuous line) and without GR (dashed line) for $a_1 = 0.05$ au (top plot) and $a_2 = 0.02$ au  (bottom plot). The outer planet has $a_2 = 0.5$ au in both cases while the initial eccentricities are 
$e_1 = 0.6$ and $e_2=0.4$. 
}
\end{figure}

If the inner planet is progressively brought closer to the star, the GR precession term becomes stronger and the value of $e_{p1}$ becomes smaller until the secular oscillations are almost fully suppressed. This behaviour is illustrated in  Fig.\ref{fig3} where  the semi--major axis of the inner planet is moved from 0.05 au (top panel) to 0.02 au (bottom panel) keeping the same the other orbital parameters i.e. the initial eccentricities of the two planets ($e_1 = 0.6$ and $e_2 =0.4$) and the semi--major axis of the outer body $a_2 = 1.5$ au. In the case with $a_1 = 0.02$ au the secular oscillations of the inner planet are almost fully damped and occur at a very fast frequency. 

This last  case with the planet very close to the star has to be considered as a tutorial test since for such close orbits tides raised by the planet on the star and viceversa,  either dynamical  \citep{ivanov2004,ivanov2007,mathis2015,ogilvie2013} or with constant phase lag \citep{jack2010, penev2014},  play an important role. They would damp the eccentricity and drive the planet to twice the pericenter distance like, for example,  in he case of  KELT 16b \citep{kelt2017}  or HATS 18b \citep{penev2016}. The scenario illustrated in Fig.\ref{fig3}, bottom plot, with $a_1 = 0.02$ au  would potentially occur just after a planet is injected on an eccentric close orbit by a planet--planet scattering event and last until the tidal evolution begins to have significant effects. The time--span during which the secular evolution without tide is a good approximation depends on how fast the tides act to damp the eccentricity. In the case of dynamical tides the tidal damping timescale can be short, of the order of some Myrs, while in the case of static tides it strongly depends on the Q--factor adopted for the planet and the star \citep{penev2014} and it typically ranges from some hundreds Myrs to a few Gyrs.

\section{Stabilizing effects of the GR precession term} 

As observed in the previous section, when the planets move closer to the star or the eccentricity of the inner planet is higher, the GR term is stronger and it causes a smaller oscillation amplitude of the eccentricity of both planets due to the $1 / (a_{p}^{5/2} (1 -e_p^2))$ dependence. 

\begin{figure}
\includegraphics[width=\columnwidth]{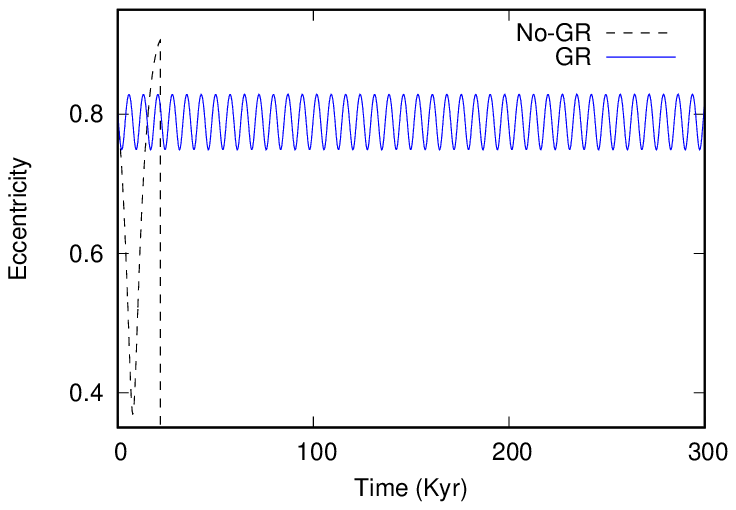}
\includegraphics[width=\columnwidth]{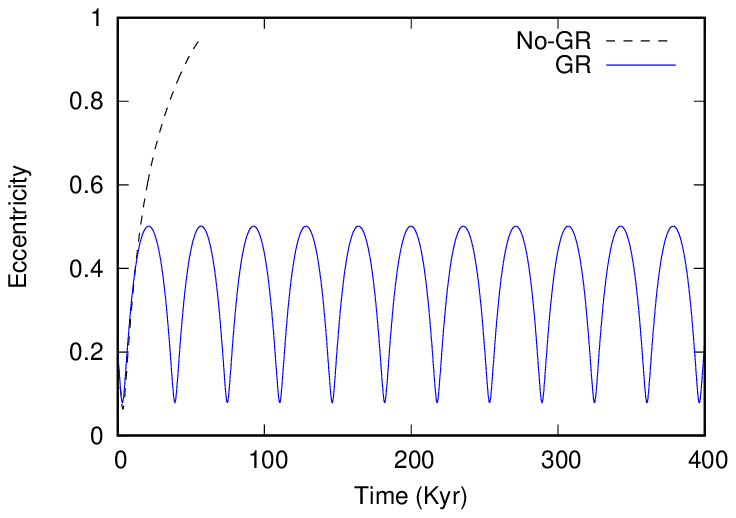}
\caption{\label{fig4} Two examples of systems which are stable with the GR contribution and unstable otherwise. The first case  (top plot) has  $a_1 = 0.05$ au,  $a_2 = 0.5$ au  and $e_1 = 0.8$ and $e_2=0.4$. The second case (bottom plot) has $a_1 = 0.1$ au,  $a_2 = 1.5$ au  and $e_1 = 0.2$ and $e_2=0.8$. 
}
\end{figure}

The reduction of $e_{p1}$  has important consequences for the long term stability of the two--planet system.  In the N--body case without GR the eccentricity oscillations induced by the secular perturbations may be large enough to cause close approaches between the planets. This may happen either directly by driving the pericenter of the outer planet within the apocenter of the inner planet or by exciting eccentricity values  high enough to enter a region of resonance overlap leading to  chaotic evolution. When GR comes into play, the oscillation amplitude of both planets is reduced and this may prevent the occurrence of  instability and close encounters. This is shown in Fig.\ref{fig4} for a case where the inner planet is very close to the star i.e.  $a_1 = 0.05$ au,  $a_2 = 0.5$ au  and $e_1 = 0.8$ and $e_2=0.4$ (top panel). The pure N--body case is unstable after a few thousands years when a close encounter between the two body occurs and the simulation is stopped. The case with the GR term shows a reduced oscillation of the eccentricity (smaller $e_{p1}$ and $e_{p2}$) preventing the planets to get too close and become unstable.  A similar situation is illustrated in the bottom panel of Fig.\ref{fig3} where in this case $a_1 = 0.1$ au,  $a_2 = 1.5$ au  and $e_1 = 0.2$ and $e_2=0.8$. In the model without GR the high eccentricity of the outer planet excites a wide secular oscillation of the eccentricity of the inner planet and a close encounter occurs when $e_1$ becomes higher than 0.9. This instability is instead suppressed when the GR term is included because $e_{p1}$ is significantly smaller and the system is then long--term stable. 

The contribution of GR allows a significant number of two planet systems to be stable in spite of their high initial eccentricities which may be the outcome of a planet--planet scattering phase. This because the reduction of the oscillation amplitudes  of the eccentricity  $e_p$ of both planets prevents close encounters which might occur in absence of the GR term. 

\section{Statistical analysis of the GR effects on the secular evolution}

The number of possible initial conditions for a system of two planets, even neglecting the initial inclinations and nodal longitudes, is very large. We have sampled the initial semi--major axis of the two planets (n.b. we cannot scale the results with the radial distance because the GR term depends on both $a_1$ and $e_1$), the initial eccentricities and their pericenter longitudes. In all the models  we fix the initial values of $\varpi_1$ and $\varpi_2$ and consider two different radial configurations. In the first one the two planets are started with $a_1 = 0.1$ au and $a_2=1.5$ au while in the second one we consider a system closer to the star and more compact  with $a_1 = 0.05$ au and $a_2=0.5$ au.
In both cases, according to 
Fig.\ref{fig21}, $g_1$ is close to the GR precession frequency and the damping of $e_{p1}$ is expected to be strong.
The case with  $a_1 = 0.05$ au differs from that with   $a_1 = 0.1$ au since both the GR term and the mutual secular perturbations between the planets are enhanced. In both cases the initial eccentricities of the two planets are sampled between 0.2 and 0.8. For each configuration we compute $e_{pi}$, $e_{fi}$, $g_1$ and $g_2$ and we compare the cases with and without GR. 

\subsection{Model 1:  weaker perturbations}

\begin{figure}
\includegraphics[width=\columnwidth]{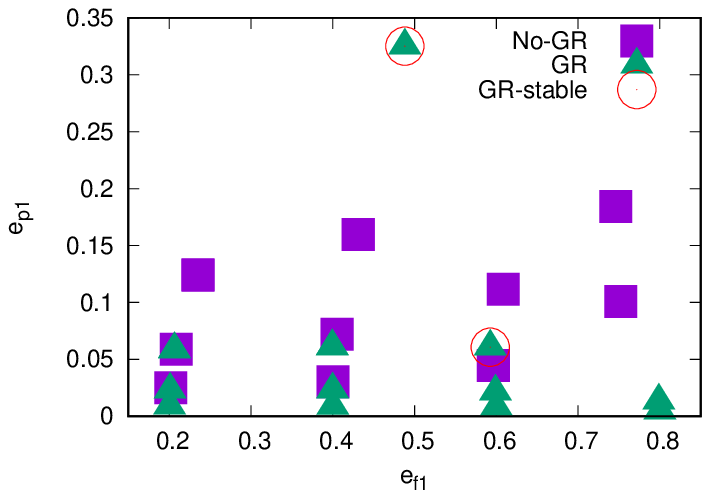}
\includegraphics[width=\columnwidth]{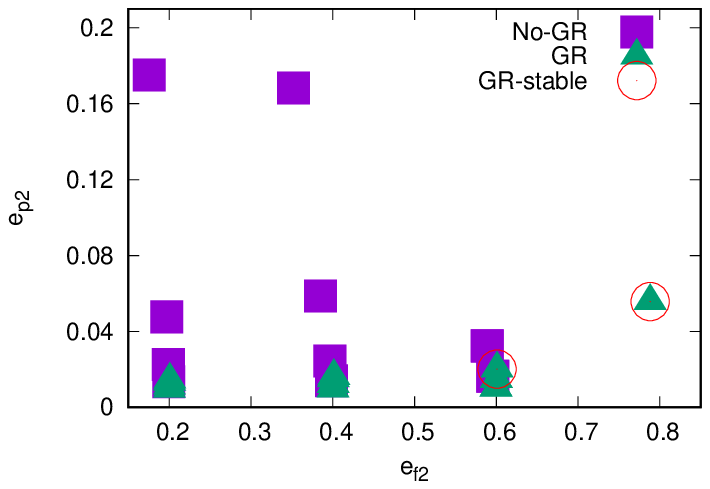}
\caption{\label{fig5} Values of $e_{p1}$ vs. $e_{f1}$ (top panel) and $e_{p2}$ vs. $e_{f2}$ (bottom panel) with GR  (green filled triangles) and without (magenta filled squares). The empty circles mark those systems which are stable only when GR is included. 
}
\end{figure}

We investigate here the evolution of systems with initial $a_1 = 0.1$au and $a_2=1.5$au.  In Fig.\ref{fig5},  upper panel, we compare the values of $e_{p1}$ and $e_{f1}$ (inner planet) with and without the contribution of GR. The values of $e_{p1}$ with GR are systematically smaller, as predicted also by the LL secular theory including the GR precession term. Also $e_{p2}$ (for the outer planet) is significantly smaller when GR is included (bottom panel) showing that the effect of the GR precession term, even if it acts on the inner planet only, is transferred to the outer planet via the mutual planetary perturbations.
The values of $e_{f}$ are not significantly affected by the GR term apart from the cases with high eccentricity where $e_{f}$ appears slightly higher when GR is included (see Fig.\ref{fig5}, top panel).

\begin{figure}
\includegraphics[width=\columnwidth]{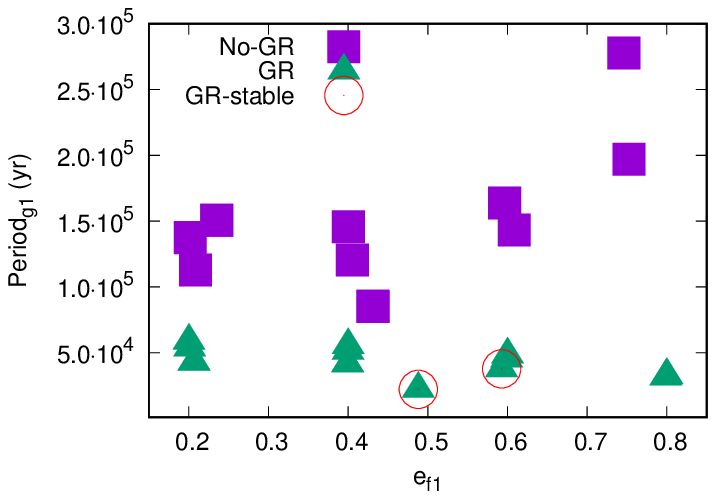}
\includegraphics[width=\columnwidth]{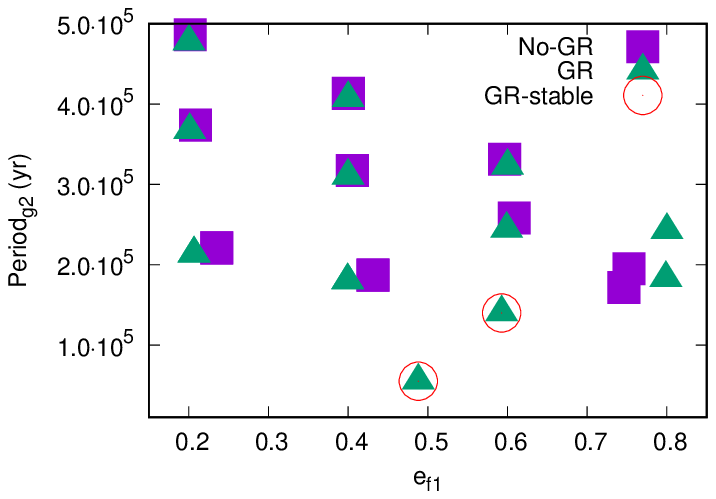}
\caption{\label{fig6} Periods of the two fundamental frequencies of the planetary systems (upper panel $g_1$, bottom panel $g_2$) as a function of $e_{f1}$. As in Fig.\ref{fig5}, the empty circles mark systems which are stable only when GR is included. 
}
\end{figure}

It is noteworthy that the cases with initial $e_{1} = 0.6, e_{2} = 0.6$, $e_{1} = 0.8,  e_{2} = 0.4, 0.6$ and $e_{1} = 0.4,  e_{2} = 0.8$,  are all unstable without GR and this explains while the growing initial trend of $e_{p1}$ vs. $e_{f1}$ seems to halt around $e_{f1} = 0.6$.  On the other hand, when GR is included both the case $e_{1} = 0.6, e_{2} = 0.6$ and $e_{1} = 0.4,  e_{2} = 0.8$ are stable and they are marked by red circles in  both Fig.\ref{fig5} and Fig.\ref{fig6}. This is an additional confirmation of the stabilizing effect of the GR term which reduces the eccentricity oscillations of the two planets. 

\begin{figure}
\includegraphics[width=\columnwidth]{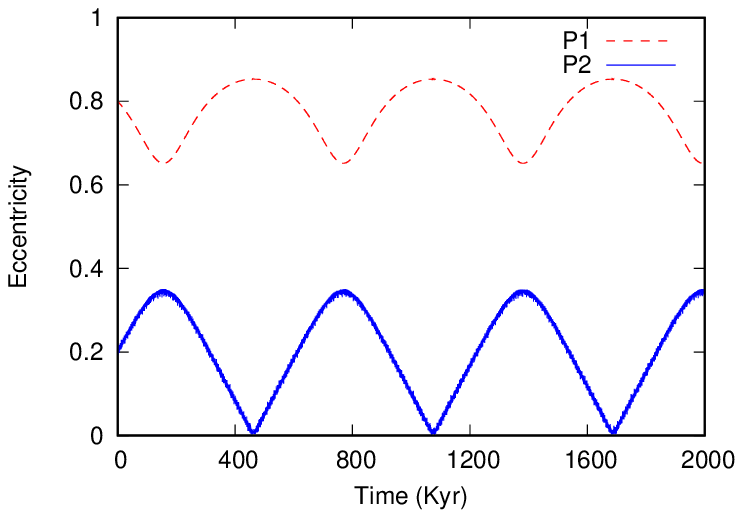}
\includegraphics[width=\columnwidth]{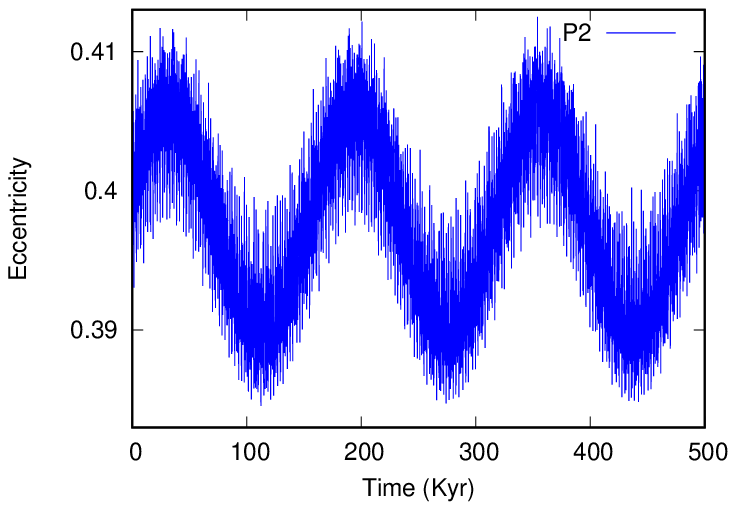}
\caption{\label{fig7} Evolution of the eccentricity for the model with initial eccentricities $e_1 =0.8$ and $e_2=0.2$ (top panel) where P1 refers to the inner planet while P2 to the outer one. In the bottom panel we show the evolution of the outer planet (P2) only when  $e_1 = 0.2$ and $e_2 = 0.4$.
}
\end{figure}

In Fig.\ref{fig6} we compare the two fundamental frequencies that, according to the LL theory, dominate the evolution of the system. As expected, one of the two frequencies, i.e. $g_1$, is significantly higher with GR and its period is smaller (top panel) and close to the GR precession period. It is remarkable  that $g_1$ depends on the eccentricity of the planets, a dependence not predicted by the LL theory that is only second order in eccentricity. In the case without GR the frequency decreases (the period grows) and this is possibly due to higher order terms in the eccentricity of the planets. On the contrary, in the case with GR the frequencies becomes higher with $e_{f1}$. This can be explained by the competition between the decreasing trend observed without GR and the increase in the relativistic precession term which is proportional to $ 1/  {(1 - e_p^2)}$. The latter wins and the frequency grows (the period decreases) with $e_{f1}$. For the second frequency $g_2$, this competition is not present because the GR term does not significantly contribute to determine it and, as a consequence,  there is no difference between the case with and without GR. In both cases the frequency grows with $e_{f1}$ (the period gets smaller) (see Fig.\ref{fig6} bottom panel) and this is, again, possibly due to higher order eccentricity terms. These same terms are also responsible of the higher complexity of the periodic evolution of both   $e_1(t)$ and $e_2(t)$ for the high eccentric cases.  In the top panel of  Fig.\ref{fig7} the curves of the eccentricity are periodic but slightly different from the sinusoidal behaviour observed in  Fig.\ref{fig1}. 
Higher order frequencies may also appear at moderate eccentricities, as illustrated on the  bottom  panel of  Fig.\ref{fig7} where the evolution of the eccentricity of the outer planet is shown in the case where initially $e_1 = 0.2$ and $e_2 = 0.4$. 

\subsection{Model 2: stronger perturbations} 

\begin{figure}
\includegraphics[width=\columnwidth]{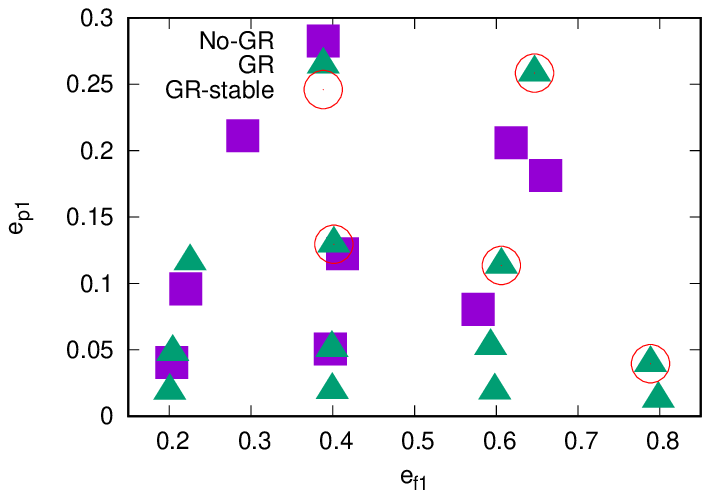}
\includegraphics[width=\columnwidth]{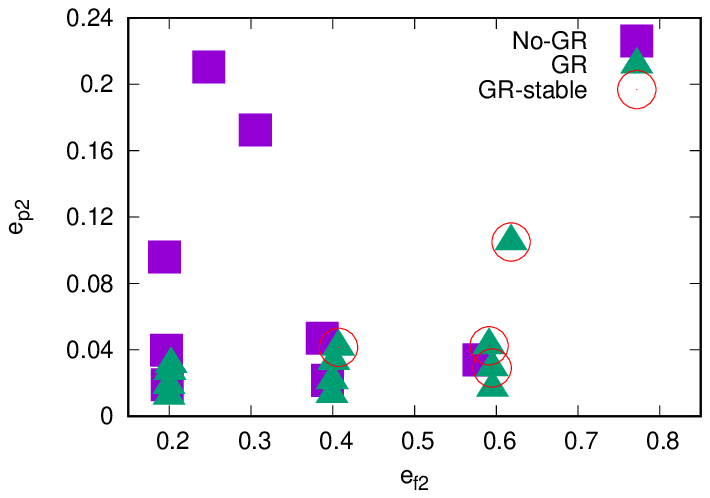}
\caption{\label{fig8} Same as in Fig.\ref{fig5} but with $a_1 = 0.05$au and $a_2 = 0.5$au.
}
\end{figure}

\begin{figure}
\includegraphics[width=\columnwidth]{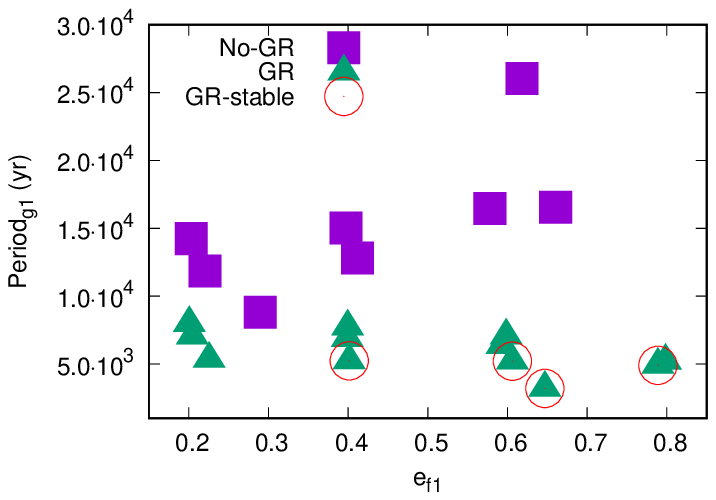}
\includegraphics[width=\columnwidth]{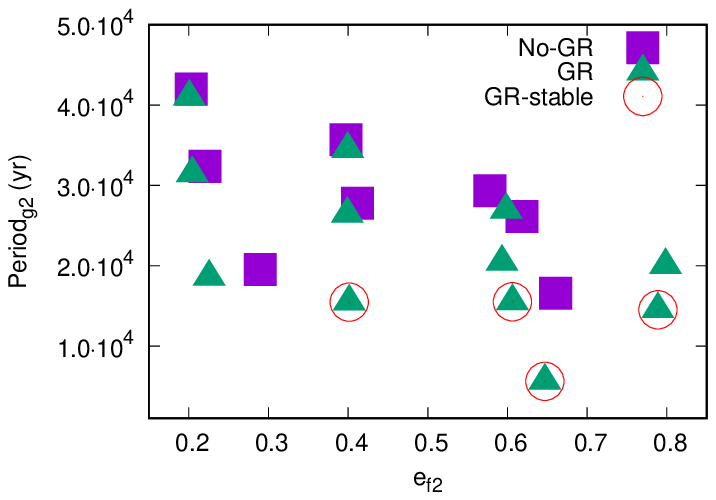}
\caption{\label{fig9} Same as Fig.\ref{fig6} but with $a_1 = 0.05$au and $a_2 = 0.5$au.
}
\end{figure}

In this second configuration we adopt $a_1 = 0.05$ au and $a_2=0.5$ au and again sample  different initial eccentricities for the two planets. The inner body is closer to the star and this implies that the GR term is stronger than in model 1. In addition, the two planets are also closer to each other than in model 1 and, therefore, also the secular interaction is stronger. This can be deduced by the higher values of $e_{p1}$ and $e_{p2}$ shown in Fig.\ref{fig8} compared to those shown in  Fig.\ref{fig5} for model 1. Even in this dynamical configuration there is a    substantial reduction of the $e_{p1}$ values when GR is included, due to the forced precession term and the increased difference between $g_1$ and $g_2$ (Fig.\ref{fig9}), as expected. Due to the stronger secular perturbations, 2 additional systems are unstable, when GR is not included, compared to the case with $a_1 = 0.1$ au and $a_2=1.5$ au. As a consequence, there are four cases which are stabilized by GR in Fig.\ref{fig8} compared to only 2 in model 1. The more packed the system is, the more effective is GR in stabilizing the system by damping  the eccentricity oscillations. 

In this more perturbed configuration, we retrieve the same behaviour observed in model 1. The values of $e_{p}$ are strongly reduced by the  GR term for both the inner and outer planets and the eccentricity has damped oscillations  around $e_{f}$. The values of $e_{f}$ do not appear to be strongly influenced by the GR term as in model 1 even if a slight increase in  $e_{f}$ is shown in   Fig.\ref{fig8} bottom panel. 

\section{Mass dependence}

In the framework of the LL theory, the matrix A linearly depends on the mass of the planets and, in a more complex way, on their semi--major axis. The ratio $a_1/a_2$  appears in the Laplace coefficients and there is an additional linear dependence of the matrix elements on it.  The GR precession term depends only weakly on the mass of the planets, through the term $(M_s + m_p)^{3/2}$, and it is inversely proportional to the semi--major axis of the planet via $1/ {a_p^{5/2}}$. While an increase in the distance between the planets may favor the GR term by increasing the secular frequencies (see Fig.\ref{fig21} and Fig.\ref{fig11}), a simultaneous change in the masses and semi--major axes would more deeply alter the balance between the secular and GR terms.  

\begin{figure}
\includegraphics[width=\columnwidth]{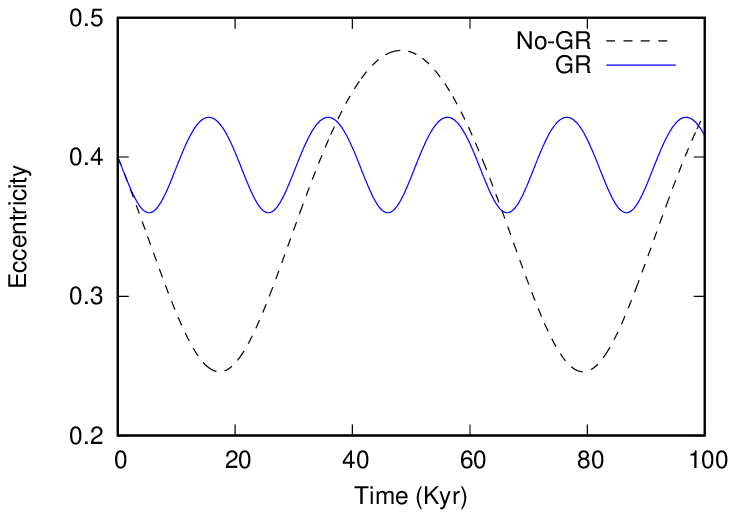}
\includegraphics[width=\columnwidth]{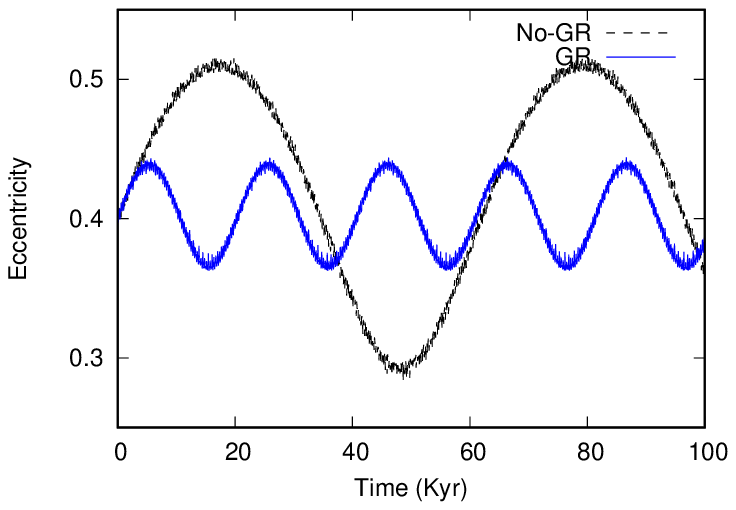}
\caption{\label{fig10} In this model the masses of the planets are 1 Jupiter mass (inner planet) and 1 Saturn mass, respectively. Their initial orbital elements are: $a_1 = 0.05$ au, $a_2=0.5$ au and $e_1=e_2=0.4$. 
The evolution of the eccentricity of the inner planet is shown in the top panel while that of the  
outer planet is shown in the in the bottom panel.
}
\end{figure}

We have performed a few additional tests where we have considered different masses for the planets. In Fig.\ref{fig10} we adopted a pair of planets with masses equal to that of Jupiter (inner body) and of Saturn (outer body). Even in this configuration there is a significant damping of $e_{p1}$ when GR is included for both planets and there is also a slight increase in the forced eccentricity of the inner planet $e_{f1}$, from 0.37 without GR to 0.4 otherwise.  

\begin{figure}
\includegraphics[width=\columnwidth]{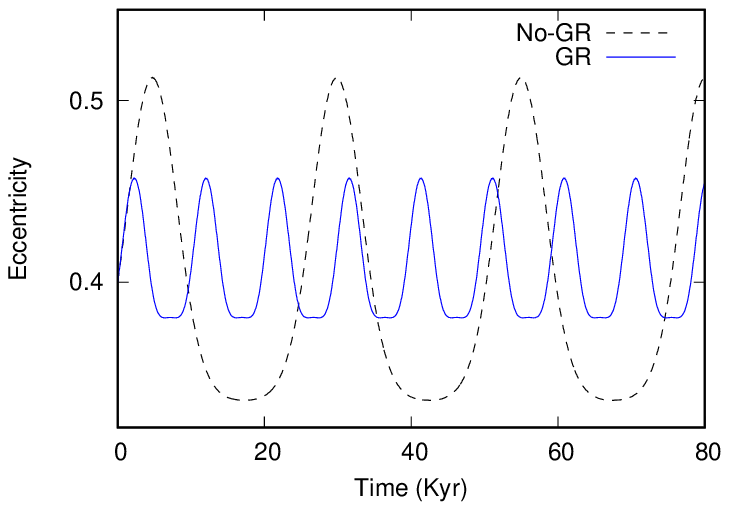}
\includegraphics[width=\columnwidth]{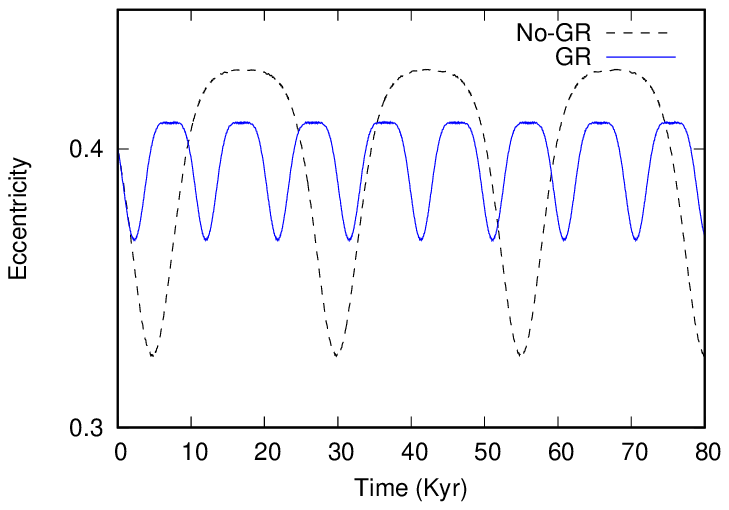}
	\caption{\label{fig10bis} In this model both the planets are super--Earths with masses equal to 5 $M_{\oplus}$. Their initial orbital elements are: $a_1 = 0.05$ au, $a_2=0.2$ au and $e_1=e_2=0.4$. The eccentricity of the inner planet is shown in the top panel while that of the outer planet in the bottom panel.
}
\end{figure}

Even super--Earth pairs are affected by GR as illustrated in Fig.\ref{fig10bis}. In this model both $e_{p1}$ and $e_{p2}$ are damped while $e_{f2}$ is slightly increased. The eccentricity oscillations are less sinusoidal but are still periodic and allows a meaningful  definition of both $e_p$ and $e_f$. This last model suggests that compact systems of super--Earth on eccentric orbits (see ad example HD 106315 or Kepler 20) may have their secular evolution significantly affected by GR which should be included when investigating their long term stability properties. 

\section{Conclusions and discussion}

Multiple planetary systems orbiting close to the star may have their secular evolution altered by the forced GR precession. We have studied a simple case where a pair of Hot/Warm Jupiters are mutually perturbing and the inner planet is also close enough to have a significant GR precession.  As a first approximation, we have adapted the LL theory by adding the GR term and recalculating the eigenvalues and eigenvectors of the modified secular matrix.
Within this framework, we have shown that the increase in the secular frequency $g_1$ due to GR leads to a larger difference between the two fundamental frequencies that, in turn, causes a reduction of the eccentricity oscillation amplitude. 

A more accurate approach based on the direct numerical integration of the equation of motion is then adopted and the data are analysed by using a parametrization based on the LL theory. We find that both the 
analytical and numerical models predict a 
significant damping of  $e_{p}$ in presence of the GR term while  $e_{f}$ is almost unchanged, except for very high initial eccentricities. In other words, the amplitude of the  oscillations of the eccentricity around $e_{f}$, which may be termed as 'forced' eccentricity, are faster and  significantly reduced.
It is interesting to notice that the effects of GR extends also to the outer planet by damping its $e_{p}$ but without affecting $e_{f}$. The GR term speeds up the frequency $g_1$, that typically dominates the evolution of the pericenter of the inner planet, while $g_2$ is significantly less affected. 
According to the prediction of the LL theory, and confirmed by numerical simulations,  it is also expected that the effects of GR on $e_p$ are stronger when the secular frequency $g_1$ (computed without the GR term)  approaches the GR precession frequency. 

The damping of $e_p$ has relevant consequences for the long term stability of the planet pair. In absence of the GR precession term, the dynamical evolution of systems on highly eccentric orbits may lead, on a short timescale, to close encounters, chaotic behaviour and disruption of the initial architecture of the system. However, if the planets are close enough to the star to have a strong GR precession, at least the inner one, the reduction of $e_p$ of both planets may stabilize the system preventing close encounters to occur and granting  long term stability. This has been found in the N--body numerical simulations and the effect is more important when the planets are closer to the star since the GR term is stronger. The latter depends also on the eccentricity of the planet and, for high initial eccentricities, it is even more effective.  

It is a complex task to derive analytical expectations on the effects of GR when also the mass of the planets are changed. The secular evolution and the GR term depend in a complex way on both $m_p$ and $a_p$ and this makes it difficult to outline a trend. However, we have shown that also less massive planets are affected by the GR precession term and a similar reduction in $e_p$ is observed, related to the increase of the difference $g_1 - g_2$ between the two fundamental
secular frequencies.  \\

{\it Acknowledgements.} We thank an anonymous referee for the helpful comments and suggestions. 
M.N. acknowledges support from
grant MEXT-KAKENHI(18H05438). 

\bibliographystyle{aa}
\bibliography{biblio}

\end{document}